\documentclass[%
 reprint,
superscriptaddress,
preprintnumbers,
nofootinbib,
 amsmath,amssymb,
 prd,
]{revtex4-2}

\usepackage{graphicx}
\usepackage{dcolumn}
\usepackage{bm}
\usepackage{booktabs}
\usepackage{xcolor}
\usepackage{slashed}
\usepackage{amssymb}
\usepackage{lipsum}
\usepackage{multirow}
\usepackage{float}
\usepackage[hidelinks]{hyperref}

\newcommand{\Neff}{N_\mathrm{eff}}
\newcommand{\Tcm}{T_\mathrm{cm}}

\begin{document}

\preprint{UCI-TR-2025-08, N3AS-25-012}

\title{Return of the Lepton Number: Sterile Neutrino Dark Matter Production and the Revival of the Shi-Fuller Mechanism}

\author{Cannon M.\ Vogel}
\email{cvogel1@uci.edu}
\affiliation{Department of Physics and Astronomy,  University of California, Irvine, California 92697-4575, USA}

\author{Helena Garc\'ia Escudero}
\email{garciaeh@uci.edu}
\affiliation{Department of Physics and Astronomy,  University of California, Irvine, California 92697-4575, USA}

\author{Julien Froustey}
\email{jfroustey@berkeley.edu}
\affiliation{Department of Physics, University of California Berkeley, Berkeley, California 94720, USA}
\affiliation{Department of Physics, University of California San Diego, La Jolla, California 92093, USA}

\author{Kevork\ N.\ Abazajian}%
 \email{kevork@uci.edu}
\affiliation{Department of Physics and Astronomy,  University of California, Irvine, California 92697-4575, USA}

\date{July 24, 2025}

\begin{abstract} 
We explore resonant production of sterile neutrino dark matter via the Shi-Fuller (SF) mechanism, revisiting its cosmological viability in light of recent results demonstrating that lepton-number asymmetries $L_\alpha \gtrsim 1$ at temperatures $T > 20\rm\,MeV$ are consistent with big bang nucleosynthesis (BBN). Using a quasiclassical Boltzmann transport calculation of the dark matter production, we compute the nonthermal phase space distributions of sterile neutrinos across a broad range of particle mass $m_s$ and mixing angle $\sin^2{(2\theta)}$ parameter space. We then evolve the resulting distributions through linear structure formation using CLASS and fit the resulting matter power spectra to thermal warm dark matter (WDM) transfer functions, enabling a direct mapping between SF models and equivalent thermal WDM particle masses $m_{\mathrm{th}}$. This allows us to reinterpret existing structure formation limits and Lyman-$\alpha$ forest preferences in the context of SF production. We find that lepton asymmetries $L \gtrsim 0.5$ at high temperatures open significant viable parameter space in the $m_s \gtrsim 10\,\mathrm{keV}$ and $\sin^2 (2\theta) \lesssim 10^{-14}$ regime, compatible with both x-ray constraints from \textit{NuSTAR} and \textit{INTEGRAL/SPI} and recent Lyman-$\alpha$ inferences of $m_{\mathrm{th}} \approx 4.1\,\mathrm{keV}$. Following lepton number evolution below 20 MeV, we also specifically show that this lepton asymmetry parameter space is compatible with BBN and cosmic microwave background constraints. We present updated constraints, a refined $m_{\mathrm{th}}$ fitting function, and power-law approximations for $L$ across the parameter space. Our results motivate future x-ray observations targeting the $\sim 20$~keV photon regime and testing of the $m_\mathrm{th} \gtrsim 10\,\mathrm{keV}$ WDM region.
\end{abstract}

\maketitle

\section{\label{sec:level1}Introduction}

Significant progress has been made in constraining a wide variety of potential candidates for the cosmological dark matter, yet the particle composition of it continues to remain a mystery. One promising class of candidates is sterile neutrinos, particles that are typically responsible for the generation of neutrino masses in theories beyond the Standard Model of particle physics. Sterile neutrino dark matter has also risen in interest because it can affect cosmological structure formation through its properties as warm to cold dark matter. In addition, it can be detected via x-ray astronomy and probed through laboratory nuclear decay experiments (for reviews, see Refs.~\cite{Boyarsky:2018tvu, Kusenko:2009up, Drewes:2016upu, Abazajian:2017tcc}).

Important clues to the dark matter problem are present in the distribution and properties of galaxies in the nearby Universe, and the clustering of gas and dark matter in the high-redshift Universe. Importantly, while $\Lambda$CDM---a model with a cosmological constant ($\Lambda$), and cold dark matter (CDM)---has been relatively successful despite its simple parametrization, there has been ongoing interest in extended theories of dark matter that modify its small-scale structure that could accommodate apparent discrepancies with $\Lambda$CDM (for a review, see Ref.~\cite{Bullock:2017xww}). One such simple modification is the introduction of a nontrivial initial velocity distribution for the particle dark matter from its initial thermal contact with the primordial plasma \cite{Blumenthal:1982mv}. Such dark matter is now referred to as \textit{thermal} warm dark matter (WDM), and its velocity dispersion results in a suppression of structure at small, sub-Galactic scales, since WDM can escape small gravitational potential wells.

WDM is one of the simplest and most studied extensions of the benchmark purely cold dark matter framework. Historically, WDM was of interest in its ability to reduce the number of dwarf galaxies in the Local Group \cite{Bode:2000gq} as well as their central densities \cite{Avila-Reese:2000nqd}. Now, there are a multitude of observations that constrain WDM: the phase-space density of galaxies \cite{Wang:2017hof,Bezrukov:2025ttd}, satellite counts \cite{Polisensky:2010rw,Cherry:2017dwu,Nadler:2019zrb}, structure in the Lyman-$\alpha$ forest \cite{Irsic:2017ixq,Palanque-Delabrouille:2019iyz}, stellar streams \cite{Banik:2019smi}, and strong lenses \cite{Keeley:2024brx}. The current strongest limits on \textit{thermal} WDM particle masses, $m_\mathrm{th}$, are from galaxy counts and strong lensing, which infer $m_\mathrm{th} > 8.4\,\mathrm{keV}$ (95\% CL) \cite{Nadler:2021dft} for galaxy counts,   $m_\mathrm{th} > 8.5\,\mathrm{keV}$ (95\% CL) from a combination of galaxy counts and strong lensing \cite{Zelko:2022tgf}. Interestingly, an analysis of Lyman-$\alpha$ forest data prefers WDM over CDM, with an inferred thermal WDM particle mass of $m_\mathrm{th} = 4.1^{+33.3}_{-1.5}\,\mathrm{keV}$ \cite{Villasenor:2022aiy}.\footnote{We have modified these $m_\mathrm{th}$ values to the more accurate thermal WDM transfer functions of Ref.~\cite{Vogel:2022odl}, and provide this correction in Eq.~\eqref{eq:th_correction}. Within their respective analyses, using prior thermal WDM transfer functions, these works infer a thermal WDM preference of $m_\mathrm{th} = 4.5^{+40.5}_{-1.4}\,\mathrm{keV}$ in the Lyman-$\alpha$ forest data \cite{Villasenor:2022aiy} and constraints of $m_\mathrm{th} > 9.7\,\mathrm{keV}$ (95\% CL) \cite{Nadler:2021dft} for galaxy counts,   $m_\mathrm{th} > 9.8\,\mathrm{keV}$ (95\% CL) from a combination of galaxy counts and strong lensing \cite{Zelko:2022tgf}.} Sterile neutrino dark matter, however, is \textit{nonthermal}, and these limits must be converted to where its particle mass produces commensurate effects \cite{Zelko:2022tgf}.

Another key observational property of sterile neutrino dark matter is its radiative decay channel, in which a sterile neutrino dominant mass eigenstate decays into a mass eigenstate that is predominantly an active neutrino and an x-ray photon with energies at half the dark matter particle mass \cite{Shrock:1974nd, Pal:1981rm}. Although this decay rate is suppressed by the small active-sterile mixing angle, $\sin^2{2\theta}$,  it increases as the fifth power of the particle mass, such that the signal from dark matter halos remains potentially observable with x-ray telescopes \cite{Abazajian:2001nj,Abazajian:2001vt}. Interest in sterile neutrino dark matter rose when Bulbul \textit{et al.}~\cite{Bulbul:2014sua} reported a high significance, $5\sigma$, detection of an x-ray line in stacked observations of 73 clusters with the MOS and PN spectrometers aboard the {\em XMM-Newton} telescope, as well as a consistent signal from the Perseus cluster of galaxies observed with the {\em Chandra} telescope. \citet{Boyarsky:2014jta} found a consistent signal from the Andromeda galaxy as well as Perseus using data from {\em XMM-Newton}. There has been significant scrutiny of the results as well as follow-up observations that see commensurate signals in other astronomical observations \cite{Abazajian:2017tcc}, and several observations do not see the line with deep exposures \cite{Ruchayskiy:2015onc,Jeltema:2015mee,Sicilian:2020glg}. Among the most significant constraints are those by \citet{Dessert:2018qih}; however, they are a factor of $\sim$20 weaker than claimed, which is acknowledged within that work, and in subsequent comments, e.g., Refs.~\cite{Boyarsky:2020hqb,Abazajian:2020unr}. Other strong limits from Foster et al.~\cite{Foster:2021ngm} do not include instrumental and on-sky lines present at 3.3 and 3.7 keV in their stated constraints. \citet{Dessert:2023fen} claim that background mismodeling led to the original line detection, though this remains disputed \cite{BulbulPrivate}.  The recently launched \textit{XRISM} Observatory will have great sensitivity to this line \cite{Dessert:2023vyl}. The \textit{XRISM} Observatory's factor of $\sim$50 greater energy resolution than \textit{Chandra} or \textit{XMM-Newton} could also potentially resolve the dark matter velocity broadening of the line \cite{Bulbul:2014sua,Lovell:2023olv}.

In this work, we evaluate sterile neutrino dark matter in light of x-ray and WDM constraints. The remaining parameter space is sensitive to observations from the \textit{NuSTAR} observatory \cite{Perez:2016tcq, Roach:2019ctw, Roach:2022lgo, Krivonos:2024yvm} and \textit{SPI}/\textit{INTEGRAL} \cite{Fischer:2022pse}. \textit{NuSTAR} has placed leading limits on sterile neutrino decay through multiple analyses: deep ``0-bounce'' observations of the Galactic Center~\cite{Perez:2016tcq}, pointed observations of the Bullet Cluster~\cite{Roach:2019ctw}, a recent long-exposure, stacked-field survey of the Milky Way halo totaling over 20~Ms~\cite{Roach:2022lgo}, and a full-mission stray-light analysis with robust background filtering spanning 11 years of data~\cite{Krivonos:2024yvm}. These studies find no evidence of monochromatic decay lines and set 95\% CL upper limits at $\sin^2(2\theta) \lesssim 10^{-12}$ in the 6--40~keV mass range. \textit{SPI}, the high-resolution spectrometer aboard \textit{INTEGRAL}, searched for decay signatures in stacked galaxy cluster fields and across the Galactic halo, similarly reporting null results and exclude lifetimes below $\sim 10^{20}$~s at 95\% CL [corresponding to $\sin^2(2\theta) \lesssim 3\times 10^{-14}$] in the $\gtrsim 40$~keV mass regime~\cite{Fischer:2022pse}. 

Many particle dark matter models propose to utilize thermal production mechanisms wherein an initial period of coupling to the standard model plasma is followed by a freeze-out which locks in the dark matter density and preserves a Fermi-Dirac phase-space distribution. Such freeze-out is also the case for early classes of WDM models \cite{Blumenthal:1982mv,Colombi:1995ze}. In contrast, WDM can be produced by a different set of mechanisms resulting in phase-space distributions that diverge from thermal for the resulting dark matter. For example, sterile neutrinos are generally produced with a nonthermal distribution, resulting from neutrino oscillations and scattering in the early Universe. In the original mechanism introduced by \citet{Dodelson:1993je} (hereafter DW), active neutrinos are suppressed in their matter-affected mixing with sterile neutrinos at high temperature, followed by production scattering rates falling with temperature, through what is often described as a freeze-in process that achieves the proper dark matter abundance. In the minimal case, the only new parameters in the DW mechanism are a (predominantly) sterile neutrino particle mass and a mixing angle with its (predominantly) active neutrino partner \cite{Dodelson:1993je}.  If there exists a lepton-number asymmetry in one or more flavors of active neutrinos in the early Universe \cite{Lattanzi:2024hnq}, during dark matter production, the neutrinos' matter-affected potential is modified, introducing the possibility of a Mikheev-Smirnov-Wolfenstein (MSW) resonance during production via the Shi-Fuller (SF) mechanism \cite{Shi:1998km}. The SF mechanism can also be embedded in more extended models, e.g. $\nu$MSM \cite{Asaka:2005pn,Laine:2008pg}. Owing to the resonance, SF allows for smaller mixing angles than DW in producing the observed relic dark matter density. While the original DW mechanism is ruled out as producing the entirety of the dark matter due to its narrow band in mass and mixing angle parameter space \cite{Horiuchi:2013noa}, adding a lepton-number asymmetry (SF) can provide more flexibility with respect to x-ray and structure formation constraints.  

The SF mechanism was thought to be highly constrained at small mixing angles since larger lepton numbers were required to produce the proper dark matter density for those mixing scales~\cite{Chu:2006ua,Kishimoto:2008ic,Cherry:2017dwu}. The lepton number or asymmetry in a neutrino flavor $\alpha$ is defined to be
\begin{equation}
    L_\alpha \equiv \frac{n_{\nu_\alpha}-\bar{n}_{\nu_\alpha}}{n_\gamma},
\label{eq:leptonnumber}
\end{equation}
with $n_{\nu_\alpha}$ ($\bar{n}_{\nu_\alpha}$) the $\alpha$ flavor (anti)neutrino number density, and $n_\gamma$ the photon number density. The limit on the lepton number from primordial nucleosynthesis and assumed equilibration of neutrino flavors from oscillations was determined to be of order $L_\alpha \lesssim 0.05$ \cite{Serpico:2005bc}. However, it was recently calculated that neutrino flavor oscillation physics locks in lepton numbers to potentially large and unequal values at early times, so that lepton numbers of order $L_\alpha \sim 10$ or larger can exist during SF dark matter production, a {\em more than 2 orders of magnitude relaxation of constraints on the SF mechanism} \cite{Froustey:2024mgf,Domcke:2025lzg}. This allows for much smaller mixing angles to produce the requisite dark matter density, with some caveats as to the particle production with such large $L_\alpha$, which we discuss below. Some initial work in this extended parameter space was performed in Ref.~\cite{Gorbunov:2025nqs}.

We also study the impacts of SF production on the inferred clustering of the dark matter with respect to WDM limits.  Although the production mechanism of the sterile neutrino WDM candidates considered here is nonthermal, a key observation we make and go on to utilize is that the resulting linear matter power spectra can be well approximated with those of thermally produced WDM models. We show that both thermal WDM and SF sterile neutrino dark matter have a similar shape to their characteristic suppression of small-scale power due to free-streaming effects. This similarity allows us to use the familiar transfer function typical of WDM to map a thermal particle mass onto each point in the SF parameter space. 

In exploring the SF mechanism to produce the dark matter consistent with all constraints, we find that the parameter space for the SF mechanism is significantly broader than that inferred in previous work, primarily due to the relaxation of the possible lepton numbers present during SF sterile neutrino dark matter production. We examine the physics of resonant production in this mechanism in this broader parameter space. Given limits from x-ray astronomy and structure formation, we present where the SF mechanism remains viable. We also present the SF parameter space commensurate with the inferred preference for WDM from Villasenor et al.~\cite{Villasenor:2022aiy}.

In the first part of this paper, we describe the particle physics of sterile neutrinos in the early Universe, particularly in the case of SF production. Next, we discuss the evolution of lepton number through and after dark matter production, showing specifically how the recent relaxation of bounds of neutrino asymmetries applies to the situations we focus on. We then describe how to implement the SF particle distribution in structure formation---Boltzmann plus gravity--- solvers, after generating the phase-space distributions (PSDs) associated with the dark matter's production. Finally, we use our dark matter production and linear structure formation pipeline to relate the SF parameter space to current limits and preferences in WDM and x-ray astronomy. We present the available window for SF to explain current observations and set the stage for future work expanding this methodology to a variety of production mechanisms.

\section{\label{sec:2}Methods}
\begin{figure*}
    \centering
    \includegraphics[width=\linewidth]{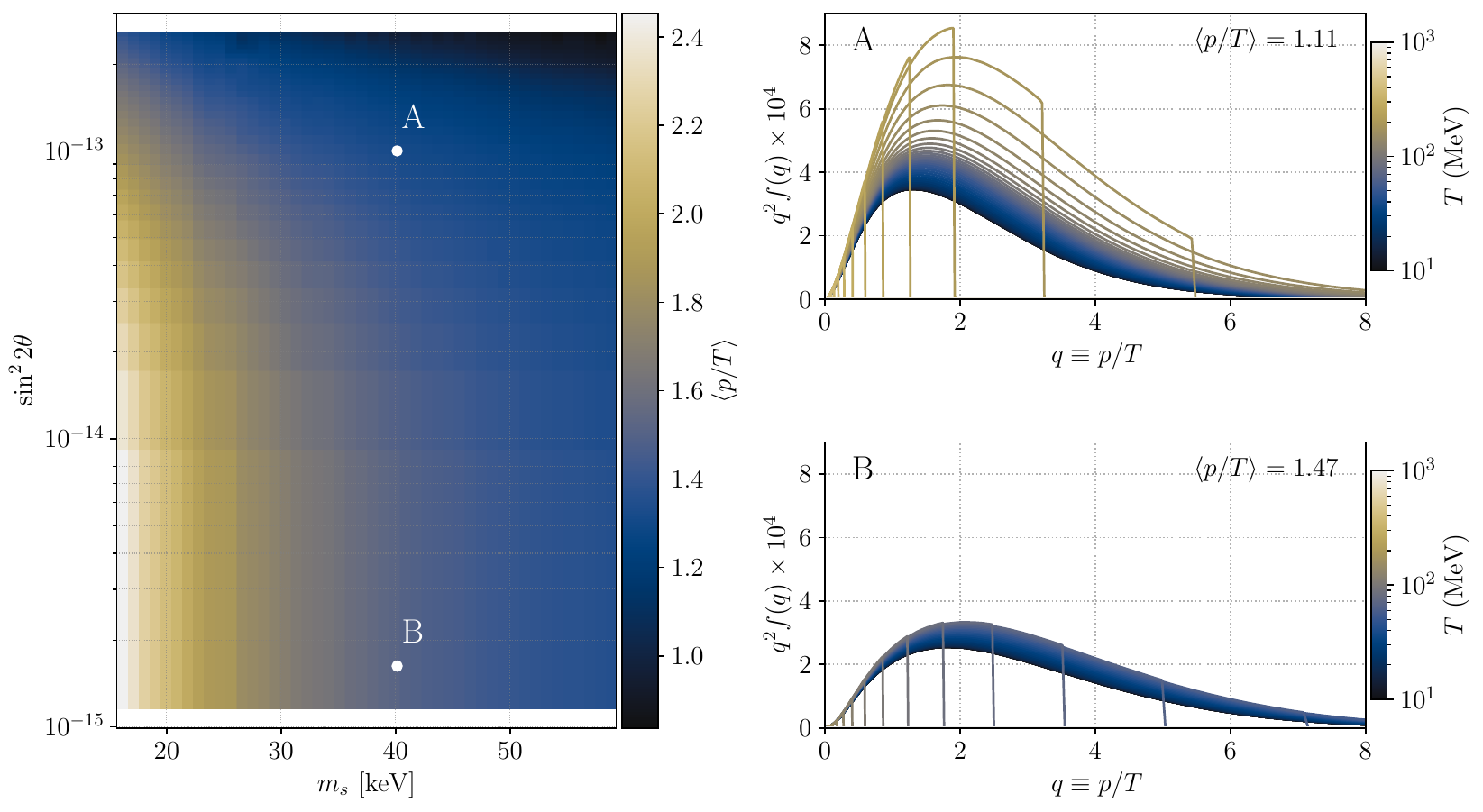}
    \caption{ Shown in the left panel is the relevant average $q_\mathrm{avg}\equiv \langle p/T\rangle$ for the region of parameter space examined in the remainder of the work. In the right panels are the PSDs' evolution as a function of temperature associated with two points on a constant sterile mass contour (corresponding to A and B in the left panel), demonstrating the sweeping of the resonance creating a hotter population at smaller mixing angles that require larger lepton asymmetries. The distributions lower their normalizations as the temperature decreases due to the production code simultaneously tracking cooling and dilution of the dark matter during production. Most dilution and cooling occurs through the QCD transition at $T=170\rm\,MeV$, which is most readily seen in case A.} 
    \label{fig:production}
\end{figure*}

\subsection{Dark matter production}
In the SF mechanism, production of sterile neutrinos occurs through scattering during active-sterile neutrino oscillations in the presence of a primordial lepton asymmetry \( L_\alpha \) that enhances production via a MSW-type resonance in the early Universe \cite{Shi:1998km}. This production mechanism can reproduce the observed dark matter density using smaller mixing angles than those required in the DW mechanism, and produces nonthermal PSDs that are colder (or ``cooler'') than thermal, {\it i.e.}, $\langle q\rangle \lesssim 3.15$, where $q\equiv p/T$. In this work, we utilize the {\tt sterile-dm} code created by Venumadhav et al. \cite{Venumadhav:2015pla}. The {\tt sterile-dm} code implements a detailed calculation of resonant sterile neutrino production in the early Universe via the SF mechanism, via a single sterile neutrino species, of variable mass and mixing angle, that mixes with the muon channel.
 
We will refer to the lepton number asymmetry in the muon flavor simply as the lepton number $L$, {\it i.e.} $L\equiv L_\mu$. The production is computed using a collision-dominated Boltzmann formalism appropriate for temperatures \( T \gtrsim 100 \, \mathrm{MeV} \), where the sterile neutrino distribution function evolves under quantum-damped, nonequilibrium dynamics, with source terms depending on the effective mixing angle in the medium, the thermal potential \( V_T \), and the asymmetry potential \( V_L \) \cite{Venumadhav:2015pla}. Between \( T \approx 100 \, \mathrm{MeV} \) and \( T \approx 20 \, \mathrm{MeV} \), very little to no dark matter production occurs in the parameter space of interest here, so that the collision-dominated Boltzmann formalism for production is sufficient. 

This implementation accounts for key plasma effects that influence sterile neutrino production. Namely, it includes a detailed calculation of the redistribution of the primordial lepton asymmetry across charged leptons, neutrinos, and hadrons through weak interactions, matched across the QCD transition using lattice QCD susceptibilities. The neutrino opacities are computed from first principles, incorporating tree-level leptonic and hadronic scattering processes, including meson decays below the QCD transition and quark-level processes above it. These parameters are used to solve the Boltzmann equation for the sterile neutrino PSDs. These PSDs are prerequisite for calculating cosmological structure formation in the linear regime, via cosmological perturbation and Boltzmann calculations. The production code outputs the PSDs across the production period and the state data of the run, providing $\Omega_{\mathrm{DM}} h^2$ and $L$ at each temperature, down to 10 MeV. While the final $\Omega_{\mathrm{DM}} h^2$ is provided as an input and taken to be the full dark matter density, the corresponding $L$ is found by {\tt sterile-dm} via a root-finding process. The code calculates scenarios for positive $L$, as production is symmetric with respect to positive or negative lepton number, with resonance occurring for antineutrinos in the case of negative $L$, with otherwise identical evolution.\footnote{The full code is available at: \href{https://github.com/ntveem/sterile-dm}{https://github.com/ntveem/sterile-dm}.}

A key feature of SF production is that the resonance condition evolves with temperature, sweeping across neutrino momentum space from low to high momenta as the Universe expands and cools, preferring production of low momenta modes in some parts of the mechanism's parameter space. Much of the physics of SF production is explored in Ref.~\cite{Kishimoto:2008ic}. From Ref.~\cite{Abazajian:2001nj}, the position of the resonance is 
\begin{eqnarray}
\label{eres}
q_{\rm res} &\approx& {\frac{\delta m^2\cos2\theta}{
{\left( 4\sqrt{2}\zeta(3)/\pi^2\right)} G_{\rm F} T^4 {|\cal{L}|} }}\\
&\approx& 0.1245 {\left({\frac{\delta m^2\cos2\theta}{1\,{\rm
keV}^2}}\right)} {\left({\frac{{10}^{-2}}{{|\cal{L}|}}}\right)}
{\left({\frac{100\,{\rm MeV}}{T}}\right)}^4 \nonumber \, ,
\end{eqnarray}
where 
\begin{equation}
\label{eq:L}
{\cal{L}^\alpha} \equiv 2 L_{\alpha} +
\sum_{\beta\neq\alpha}{L_{\beta}}\, .
\end{equation}
For simplicity, we assume that the sterile neutrino mixes exclusively with the (muon) neutrino flavor, so that $\alpha = \mu$. $G_F$ is the Fermi constant, and $\delta m^2$ is the difference between the predominantly sterile and active mass eigenstates, $\delta m^2 \equiv m_2^2 -m_1^2$. At a fixed sterile neutrino mass and temperature, increasing the lepton asymmetry shifts the resonance to lower momenta. Through production, as the Universe cools and lepton number is destroyed by dark matter production, a dynamic sweeping effect leads to a distinct imprint on the sterile neutrino PSDs. For sufficiently small asymmetries, the resonance is inefficient, resulting in a warm spectrum similar to DW production. As the asymmetry increases, the resonance becomes more efficient and spans a wider momentum range, producing a colder distribution peaked at low $q$. However, for very large asymmetries, the resonance is delayed to lower temperatures where neutrino collisions become inefficient, and the production of sterile neutrinos during resonance sweep may be truncated before populating higher-momentum modes. This results in a mixed spectrum with a cold core and a warm tail. 

These combined effects during production are shown in Fig.\ref{fig:production}, where two PSDs with the same sterile neutrino mass exhibit different average momenta. The case with lower mixing—and thus higher lepton asymmetry—is visibly warmer. Therefore, the PSDs go from warm near the DW mechanism (small $L$), to colder at intermediate $L$, to warmer at the highest $L$ values. For a fixed sterile neutrino dark matter particle mass, this warm–cold–warm evolution in the PSD is the key result from how lepton asymmetry affects production in the SF mechanism. 

There are limitations to the calculation of production that we employ, which can be placed in two categories: first, there are approximations which break down as $L$ approaches and surpasses unity; and, second, the approximations of the two-neutrino mixing treatment of production via a quasiclassical Boltzmann evolution. Regarding the first category, while exploring where the SF mechanism can produce the requisite dark matter density, we therefore push the code to larger sterile neutrino masses and smaller mixing angles, extending to large $L \sim 10 $ lepton asymmetries. In the production calculations, Fermi blocking due to large $L$ is present in the quasiclassical Boltzmann evolution equations. However, the presence of large $L$ is not included in the calculations of the scattering cross sections. Therefore, our work at $L \gtrsim 1$ is approximate and meant to be a first exploration of this parameter space. 

The second category of approximations includes the assumption of one flavor carrying all the asymmetry, collisional decoherence treated by a simple damping term, and linear perturbative redistribution of the asymmetry in the plasma. Moreover, since lepton numbers can be redistributed due to scattering effects as well as neutrino oscillations, a more accurate calculation would solve the full quantum kinetic equations in four flavors: the sterile neutrino as well as the three active flavors, which can all carry lepton number and could all mix with each other. Such conditions can produce multiple MSW level crossings and, in the case of multiple sterile neutrinos, conditions could exist for lepton number generation \cite{Foot:1995bm,Shi:1996ic,Abazajian:2004aj}. The effects of large $L$ also complicate this second category of approximations. We leave these more thorough calculations for future work.  Given these caveats at $L\gtrsim 1$, our analysis below provides the contours where $L$ becomes large in order to highlight the region where a more detailed approach is potentially necessary. 

\begin{figure}[t!]
    \centering
    \includegraphics[width=\columnwidth]{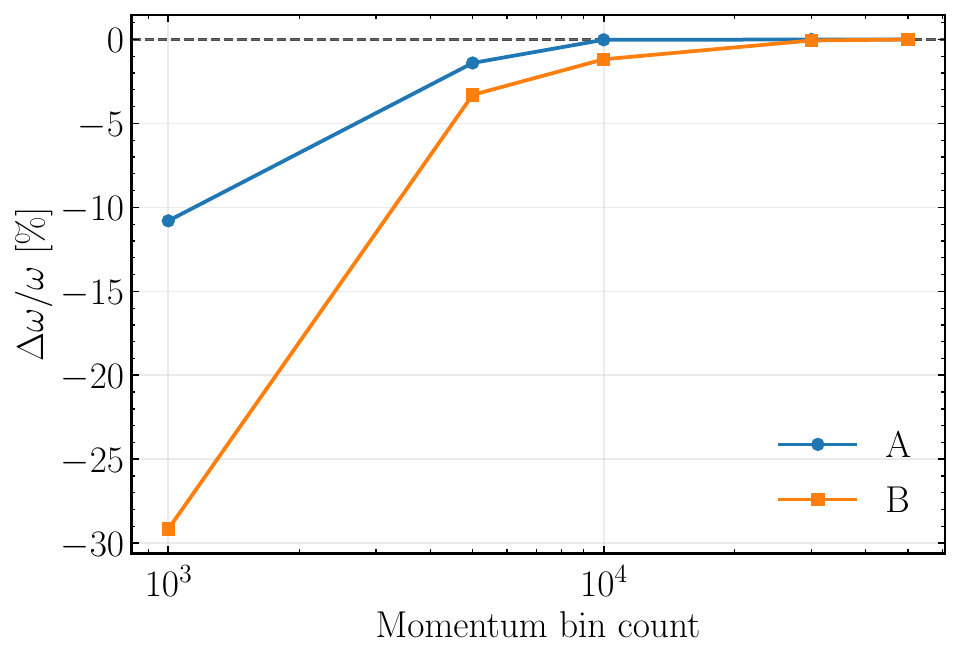}
    \caption{The blue (orange) curve shows the percentage error in the dark matter relic density, $\omega \equiv \Omega_\mathrm{DM} h^2$, as a function of momentum bin number in the {\tt sterile-dm} code for the sterile neutrino parameters associated with point A (B) in Fig.~\ref{fig:production}.} 
    \label{fig:omega-p}
\end{figure}

In addition, we note that previous work has found that increasing the momentum bin count in {\tt sterile-dm} is increasingly important as the analysis moves to a higher lepton number~\cite{Bodeker:2020hbo}. It appears that part of the resonance is missed if the momentum resolution is low, resulting in less dark matter production. Practically, this manifests as a larger lepton number identified by the code if using an insufficient resolution. In order to mitigate this issue, we analyzed different regions in parameter space before completing our analysis, finding in agreement with the authors of Ref.~\cite{Bodeker:2020hbo} that 30,000 momentum bins is generally sufficient; however, for regions near nonresonant production, we use fewer bins (down to 5,000) and for very large lepton number regions we use more (up to 50,000) to balance computational efficiency with the maintenance of accuracy at the single-percent level ($\approx 1\%$). Figure~\ref{fig:omega-p} shows the results of our analysis for the two points A and B from Fig.~\ref{fig:production}. The results demonstrate the necessity for higher resolution at relatively large lepton number (smaller mixing, such as point B) but the lesser importance in the low lepton number (larger mixing, such as point A) regime.

\subsection{Lepton number evolution}
\label{sec:lepton}
We explore lepton number evolution in two distinct periods: first, as it evolves during dark matter production, and second, after dark matter production is completed, and synchronous neutrino oscillations occur. These epochs are separated by the temperature of $T\approx 20\rm\,MeV$.

In other works, it is common to give $L$ as a ratio to the temperature or entropy in order to more easily implement neutrino functions in the Boltzmann transport equations. We provide here conversions to these common forms to assist in understanding our work in the context of the literature. First is the relation between number density asymmetry normalized by temperature cubed,
\begin{equation}
    \eta_\alpha \equiv \frac{n_{\nu_\alpha}-\bar{n}_{\nu_\alpha}}{T^3} = \frac{2\zeta(3)}{\pi^2} L_\alpha \approx \frac{L_\alpha}{4}\, .
\end{equation}
And, second, is the scaled lepton number density asymmetry normalized to entropy density $s$,
\begin{equation}
    L_6 \equiv 10^6\,\left(\frac{n_{\nu_\alpha}-\bar{n}_{\nu_\alpha}}{s}\right) = 10^6\,\left(\frac{45\eta_\alpha}{2\pi^2g_{*s}} \right) \approx 5\times 10^4\,  L_\alpha\,. \nonumber 
 \end{equation}

In order to best utilize recent findings, we will treat the lepton asymmetry as being unconstrained at the order $L \sim 10$, based on Froustey and Pitrou (FP)~\cite{Froustey:2024mgf} and in contrast to previous treatments of SF production that considered $L$ as evolving through a synchronized neutrino oscillation equalization process, and thus placed a constraint on models utilizing $L>0.05$ \cite{Dolgov:2002ab,Abazajian:2002qx,Wong:2002fa}. FP revisit cosmological constraints on primordial lepton asymmetries using a full three-flavor, multimomentum neutrino transport code that solves the quantum kinetic equations with exact collision terms and self-consistently couples to a big bang nucleosynthesis (BBN) solver \cite{Froustey:2024mgf}. The authors demonstrate that in order to satisfy BBN bounds, one need not have asymmetries of order $L\sim 0.1$, or smaller, but that in scenarios with large or flavor-imbalanced initial asymmetries, initially very large asymmetries at $T\gg 1 \,\mathrm{MeV}$ can equilibrate to small, unconstrained $L_\alpha$ when entering the BBN era. We illustrate these results specifically for the parameter space we are interested in in Sec.~\ref{subsubsec:L_evol_below20}.

Finally, we note that the lepton potential in sterile neutrino dark matter production depends on the total lepton number in an asymmetric way, with a factor of 2 greater dependence on the lepton number in the species with which the sterile neutrino mixes. The lepton potential is \cite{Abazajian:2001nj}
\begin{equation}
\label{eq:vl}
V_L = \frac{2 \sqrt{2} \zeta (3)}{\pi^2}\,G_{\rm F} T^3 \left({\cal
L}^\alpha \pm \frac{\eta}{4}\right)\, ,
\end{equation}
where $\eta$ is the baryon asymmetry, $\mathcal{L}^\alpha$ is given in Eq.~\eqref{eq:L}, and the neutrino with which the predominantly sterile species mixes is $\nu_\alpha$, where in our case $\alpha = \mu$.

\subsubsection{Cosmic lepton number evolution above $T\approx 20\rm\,MeV$}
\label{subsubsec:L_evol_above20}

Our calculations start with asymmetries well-prior to all dark matter production at $T = 10\,\mathrm{GeV}$, and end after all dark matter production, at $T=10\,\mathrm{MeV}$. It was found in initial work \cite{Dolgov:2002ab,Abazajian:2002qx,Wong:2002fa} and in FP that large asymmetries are largely frozen in evolution at $T\gg 10\,\mathrm{MeV}$ due to any active-neutrino mixing. However, during sterile neutrino dark matter production, $L$ is affected by conversion to the sterile neutrinos' ``lepton number,'' redistribution of any initial neutrino asymmetry into charged lepton and hadronic asymmetries, as well as redistribution into the other neutrino flavor asymmetries \cite{Venumadhav:2015pla}.

The redistribution of the lepton number during dark matter production is computed self-consistently in {\tt sterile-dm} by solving a set of coupled linear equations that enforce the chemical and kinetic equilibrium among the various components in the plasma \cite{Venumadhav:2015pla}. These constraint equations arise from the relation between the lepton number asymmetries and the corresponding chemical potential and number-density susceptibility of each lepton (both charged and neutral), as well as relations describing charge and baryon number conservation. The temperature dependent susceptibilities are carefully constructed across three epochs: using perturbative QCD at temperatures well above the quark-hadron transition ($T\gtrsim 300 \, \mathrm{MeV}$), lattice QCD near the transition ($150 \, \mathrm{MeV} \lesssim T \lesssim 300 \, \mathrm{MeV}$), and a hadron resonance gas model well below the transition ($T \lesssim 150 \, \mathrm{MeV}$), with care taken to smoothly match each regime. This framework captures how the initial neutrino lepton asymmetry in one flavor is dynamically redistributed into both charged lepton asymmetries, as well as into the hadronic sector. 

As an example, $\nu_\mu +\mu^+ \rightleftharpoons \pi^+$, which is faster than the Hubble rate in the dark matter production period, is one of the primary channels of this redistribution at temperatures near peak production. Redistribution between leptonic flavors becomes more dominant at lower temperatures below the quark-hadron transition. The result of redistribution across the dark matter production period is a $\simeq 10 \%$ correction in the asymmetry potential \cite{Venumadhav:2015pla}. However, as shown below, the corresponding change in $L$ across the same period is quite large, up to a factor of 10 in the examined region of parameter space.  

\begin{figure}[t!]
    \centering
    \includegraphics[width=\columnwidth]{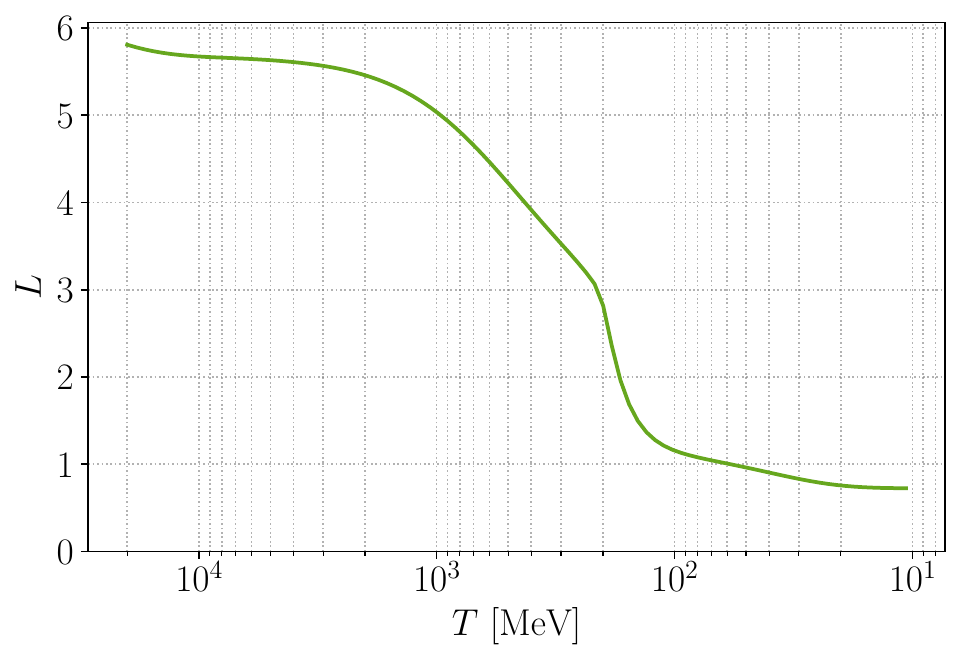}
    \caption{Evolution of the muon lepton asymmetry above $20 \, \mathrm{MeV}$, as tracked by {\tt sterile-dm}. This example is for a sterile neutrino with a particle mass of 40 keV and a mixing angle of $\sin^2{(2\theta)}=1.6 \times 10^{-15}$, corresponding to point B in Fig.~\ref{fig:production}. The lepton number is seen to change by roughly a factor of 8 in this example.} 
    \label{fig:reduceL_factor8}
\end{figure}

In Fig.~\ref{fig:reduceL_factor8}, we show an example case of the evolution of $L$ for the successful production of the dark matter for a sterile neutrino with a particle mass of 40 keV and a mixing angle of $\sin^2{(2\theta)}=1.6 \times 10^{-15}$, corresponding to point B in Fig.~\ref{fig:production}. As can be seen in the figure, production conversion and redistribution of $L$ leads to its reduction in magnitude by a factor of greater than 5.

\subsubsection{Cosmic lepton number evolution below $T\approx 20\rm\,MeV$ and cosmological constraints}
\label{subsubsec:L_evol_below20}

Below $T \sim 20 \, \mathrm{MeV}$, synchronized neutrino oscillations redistribute the asymmetry between the electron, muon, and tau neutrino flavors~\cite{Dolgov:2002ab,Abazajian:2002qx,Wong:2002fa}. Although the general trend is an approximate equilibration of the individual lepton numbers $L_\alpha$ to their average $L_\mathrm{av} = (\sum_\alpha L_\alpha)/3$, recent studies~\cite{Froustey:2021azz,Froustey:2024mgf,Domcke:2025lzg} have shown that the use of modern values of the mixing parameters and the full collision term significantly affect the details of neutrino asymmetry evolution. Lepton asymmetries can be constrained via their effect on cosmological observables, namely, the effective number of relativistic species $\Neff$ and the primordial abundances of helium-4 $Y_p$ and deuterium $\mathrm{D/H}$ (see, e.g., Refs.~\cite{Simha:2008mt,Shimon:2010ug,Oldengott:2017tzj}). Given the current experimental data on these various quantities, the strongest bounds come from BBN spectroscopic measurements (at the single-percent level---$\approx 1\%$), which tightly constrain the electron neutrino asymmetry $L_e$ \emph{at the BBN epoch}. Although it had been recognized before that specific configurations, with potentially large individual $L_\alpha$, could lead to small values of the final $L_e$ (see e.g.,~\cite{Barenboim:2016shh}), FP have shown that a large domain of the parameter space with $L \gtrsim 1$ at $T \approx 25 \, \mathrm{MeV}$ is allowed by experimental data~\cite{Froustey:2024mgf}.

\begin{figure*}
    \centering
    \includegraphics[width=\textwidth]{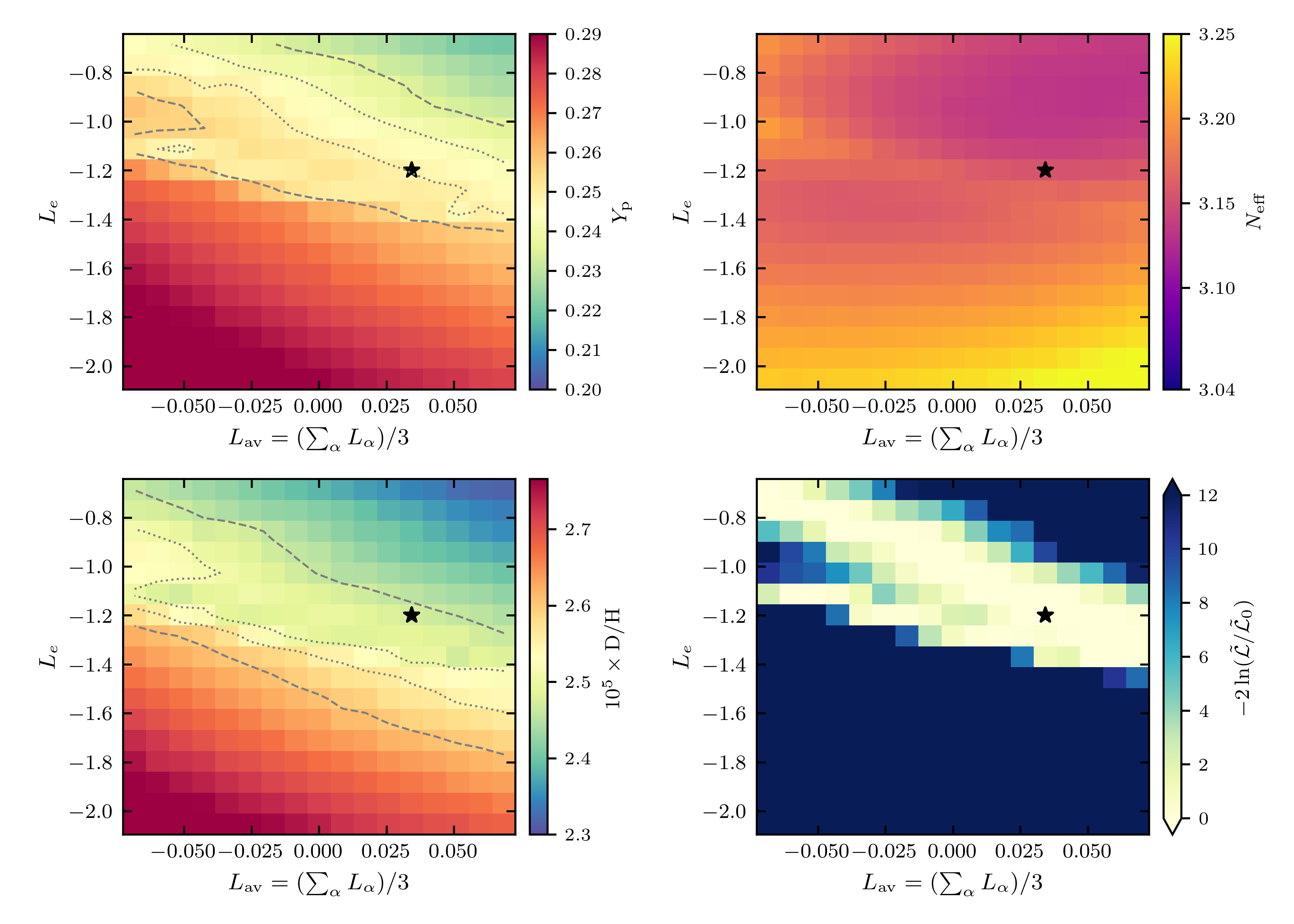}
    \caption{Shown here are outputs of the numerical solution of the neutrino quantum kinetic equations and self-consistent BBN calculations for initial asymmetry configurations at $T = 20 \, \mathrm{MeV}$ with $L_\mu \approx 0.68$, varying $L_e$ and the average lepton number $L_\mathrm{av}$. Left panels show helium-4 (top) and deuterium (bottom) abundances for a baryon density $\Omega_b h^2 = 0.0224$ (preferred value from \emph{Planck}~\cite{Planck:2018vyg}). The dotted (dashed) lines are the 68 \% (95 \%) confidence levels from the spectroscopic measurements of $Y_p$~\cite{Aver:2020fon} and $\mathrm{D/H}$~\cite{Cooke:2017cwo,Kislitsyn:2024jvk,Guarneri:2024qxi}. Top right panel shows the effective number of relativistic species $\Neff$, which is always larger than the standard value $3.044$ at zero asymmetries~\cite{Akita:2020szl,Froustey:2020mcq,Bennett:2020zkv,Drewes:2024wbw}. Bottom right panel is the likelihood based on CMB and BBN measurements (see details in~\cite{Froustey:2024mgf}), compared to the likelihood $\tilde{\mathcal{L}}_0$ of the symmetric baseline $L_{e,\mu,\tau} = 0$.}
    \label{fig:grid_asymmetries}
\end{figure*}

In what follows, we show specifically that many asymmetry configurations with a large $L_\mu$ (which is the case considered here for dark matter production) are compatible with experimental constraints. For that purpose, we apply the numerical code used in Ref.~\cite{Froustey:2024mgf} (see also Ref.~\cite{Froustey:2021azz}) to a range of asymmetry configurations at $T = 20 \, \mathrm{MeV}$, with a fixed muon neutrino asymmetry $\eta_\mu = 1/6$, that is, $L_\mu \approx 0.68$. We assume a normal ordering of neutrino masses and use the mixing parameters from~\cite{ParticleDataGroup:2024cfk}. We show in Fig.~\ref{fig:grid_asymmetries} the various cosmological observables ($Y_p$, $\mathrm{D/H}$ and $\Neff$) obtained for each initial configuration. For primordial nucleosynthesis, the (anti)neutrino distributions obtained from the solution of the quantum kinetic equations are used as input in the BBN code \texttt{PRIMAT}~\cite{Pitrou:2018cgg}; we show the resulting abundances obtained for the preferred value of the baryon density from \emph{Planck} ($\Omega_b h^2 = 0.0224$ \cite{Planck:2018vyg}). On the bottom right panel, we compute the likelihood $\tilde{\mathcal{L}}$ of each point, as a combination of Gaussian likelihoods for each BBN spectroscopic measurement (helium-4~\cite{Aver:2020fon}, deuterium~\cite{Cooke:2017cwo,Kislitsyn:2024jvk,Guarneri:2024qxi}) and the likelihood associated to the combination of CMB data described in Ref.~\cite{Froustey:2024mgf}. We then plot the ratio of this likelihood to that of the lepton-symmetric configuration $(L_e, L_\mu, L_\tau)=(0,0,0)$. For reference, the point of highest likelihood in this limited region of the parameter space is actually preferred to the ``standard model'' origin, with $- 2 \ln(\tilde{\mathcal{L}}_\mathrm{max}/\tilde{\mathcal{L}}_0) \approx - 0.94$. For visualization purposes, we represent the confidence levels of the spectroscopic measurements of primordial abundances on the left panels of Fig.~\ref{fig:grid_asymmetries}. The shift between the preferred regions for $Y_p$ and $\mathrm{D/H}$ stems from the “deuterium tension” in \texttt{PRIMAT}~\cite{Pitrou:2020etk}, which will not be resolved or confirmed until better determinations of the deuterium burning nuclear rates are available~\cite{Pitrou:2021vqr}.

\begin{figure}[!ht]
    \centering
    \includegraphics[width=\columnwidth]{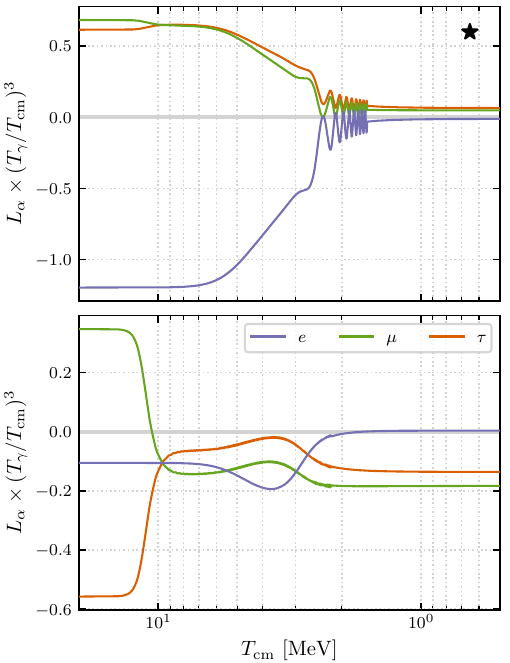}
    \caption{Shown is the evolution of the (anti)neutrino flavor asymmetries below $20 \, \mathrm{MeV}$ in two examples. The top panel shows the configuration identified with a black star in Fig.~\ref{fig:grid_asymmetries}. The bottom panel shows a configuration outside of the parameter space covered in Fig.~\ref{fig:grid_asymmetries}, but which still results in cosmological observables compatible with observations.}
    \label{fig:example_evolution} 
\end{figure}

Figure~\ref{fig:example_evolution} shows the evolution of neutrino asymmetries below $T = 20 \, \mathrm{MeV}$ for two configurations: on the top panel, a preferred point from the parameter space explored in Fig.~\ref{fig:grid_asymmetries} (shown with a black star), and on the bottom panel a configuration outside of the range of Fig.~\ref{fig:grid_asymmetries}. Since the temperature range shown on Fig.~\ref{fig:example_evolution} covers neutrino decoupling and electron-positron annihilations, we need to distinguish between the photon temperature $T_\gamma$ and the comoving temperature $T_\mathrm{cm} \propto a^{-1}$ (with $a$ the scale factor), which would be the neutrino temperature in the instantaneous neutrino decoupling approximation. Because of the variation of $T_\gamma/\Tcm$, the lepton number $L_\alpha$ defined in Eq.~\eqref{eq:L} is not a comoving quantity, which is why we multiply it by $(T_\gamma/\Tcm)^3$ in Fig.~\ref{fig:example_evolution} (see also the discussion in Ref.~\cite{Froustey:2024mgf}).

These two examples result in a very small electron neutrino asymmetry at the BBN epoch, which is the main parameter determining BBN abundances. We note that future Simons Observatory and CMB-S4-like observations will be able to constrain $\Neff$ at much better levels, and measure $Y_p$ at a level of precision comparable to current spectroscopic measurements~\cite{SimonsObservatory:2018koc, SimonsObservatory:2025wwn, CMB-S4:2016ple}, thus being able to identify or constrain these required $L$ for the SF mechanism.
    
With the results of Figs.~\ref{fig:grid_asymmetries} and \ref{fig:example_evolution}, we highlight the variety of currently allowed asymmetry configurations for a given muon lepton number $L_\mu$. In the following, we then consider the lepton number $L$ which enters dark matter production as being unconstrained at the order $L \lesssim 10$ (recalling that it decreases by a factor of at least 5 between $T=10\rm\,GeV$ and $T = 20 \, \mathrm{MeV}$).

\subsection{Cosmological structure}

WDM affects cosmological structure formation due to free streaming of the dark matter. The nature of the free streaming is unique in the case of thermal WDM, but widely varied in the case of nonthermal WDM such as DW and SF sterile neutrino dark matter. The effects of free streaming are imprinted on the distribution of matter and affect the linear matter power spectrum, and related nonlinear cosmological observables. To produce the linear matter power spectra for each generated SF PSD, we used the CLASS Boltzmann code~\cite{Blas:2011rf}. The benchmark parameters for $\Lambda$CDM are from the \emph{Planck} 2018 minimal case of one massive neutrino species with $\Sigma m_\nu = 0.06$ eV, modeled as a non-CDM dark matter component \cite{Planck:2018vyg}. Specifically, we adopt the cosmological parameters that are the best estimate parameters from the TT,TE,EE+lowE+lensing column in Table 2 of Ref.~\cite{Planck:2018vyg}. To produce the matter power spectra for the SF cases, we provide the PSDs and particle masses to CLASS. CLASS rescales the PSDs to match its normalization conventions. 

We make the assumption that all lepton asymmetries are small by the time of structure formation. That is, they can correspond to lepton number configurations such as in the top panel of Fig.~\ref{fig:example_evolution}, so that the asymmetries are small enough that the structure formation is mostly sensitive to the sterile neutrino dark matter component, and not the small remaining active asymmetries. However, in general, SF mechanisms will likely be connected with nontrivial signals in $L_\alpha$ and $\Neff$ in the BBN and CMB eras. 

While using CLASS, we examined multiple numerical tolerances and integration methods to ensure the best results for each region of parameter space, in terms of numerical stability of the results as well as optimal numerical speed; our specific implementation is available in the public pipeline code.\footnote{The code is available at \href{https://github.com/garciaeh77/GZA}{https://github.com/garciaeh77/GZA}.}

\begin{figure}
    \centering
    \includegraphics[width=\linewidth]{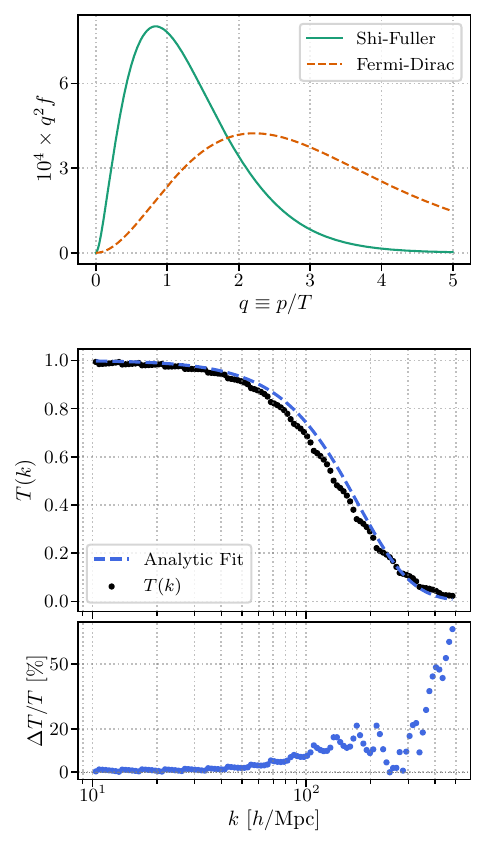}
    \caption{The top panel shows a comparison of the PSD of a 37 keV sterile neutrino with $\sin^2{(2\theta)}=1.5 \times 10^{-14}$, as well as the corresponding Fermi-Dirac distribution, demonstrating the population of lower momentum modes. The middle panel shows the corresponding relative transfer function from CLASS, compared to the thermal mass fit. The bottom panel shows the resulting residuals from the relative transfer function fitting, demonstrating the validity of our association between SF sterile neutrinos and a $T(k)=1/2$ suppression scale associated with thermal WDM at the $\sim 20\%$ level.}
    \label{fig:three_panel}
\end{figure}

\begin{figure*}
    \centering
    \includegraphics[width=\linewidth]{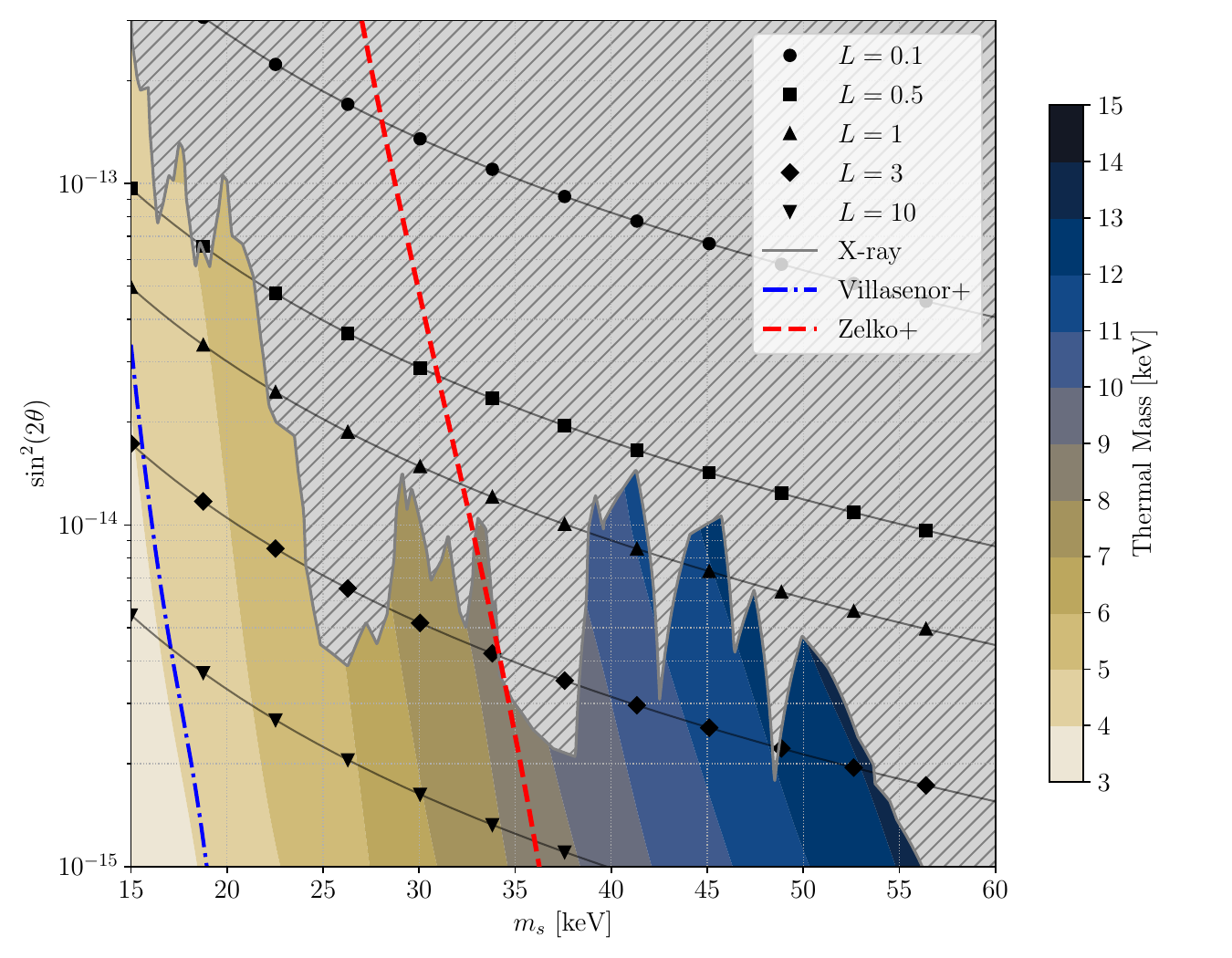}
    \caption{Shown is the full sterile neutrino dark matter parameter space under SF production, with the initial $L$ values given at $T=10\rm\,GeV$ in order to achieve the observed density of the dark matter. These $L$ values are all reduced by a factor of 5 to 10 by $T=10\,\rm MeV$ from dark matter production and dilution (see Sec.~\ref{subsubsec:L_evol_above20} and Fig.~\ref{fig:reduceL_factor8}). The $L$ is further reduced by synchronized oscillations prior to BBN, and the initial $L$ values shown are consistent with BBN and CMB observations today (see Sec.~\ref{subsubsec:L_evol_below20}). The blue dot-dashed line is the best fit the thermal WDM particle mass from Villasenor+~\cite{Villasenor:2022aiy}. The full region shown is consistent with the $1\sigma$ thermal WDM particle mass region from Villasenor+. The dashed red contour is the thermal WDM particle mass lower bound from the Zelko+~\cite{Zelko:2022tgf}, disfavoring the region left of the line. The gray cross-hatched x-ray exclusion region is the combination of \textit{NuSTAR} and \textit{SPI}/\textit{INTEGRAL} at 95\% CL upper limits \cite{Perez:2016tcq, Roach:2019ctw, Roach:2022lgo, Krivonos:2024yvm, Fischer:2022pse}.}
    \label{fig:thermal_mass_plot}
\end{figure*}

The resulting matter power spectra were converted into relative transfer functions 
\begin{equation}
\label{eq: transfer}
    T(k) \equiv \sqrt{P_X(k)/P_{\mathrm{\Lambda CDM}}(k)} \, ,
\end{equation}
and an example is shown in Fig.~\ref{fig:three_panel}. We fit the SF transfer functions with thermal WDM transfer functions, where the thermal WDM particle mass is the free parameter for each sterile neutrino particle mass and mixing angle point. As shown, the residuals between the closest-fit thermal WDM transfer functions and those of the SF-produced sterile neutrinos are relatively small ($\lesssim 20\%$) for the suppression scale of 1/2. The $\lesssim 20\%$ accuracy holds in our fit across the examined parameter space, despite many points having PSDs that are highly nonthermal.  In order to properly convert most of the prior literature's inferred thermal WDM particle masses, $m_\mathrm{th,old}$ \cite{Viel:2005qj} (more accurate at $m_\mathrm{th} \equiv m_\mathrm{th,new} \lesssim 5\,\mathrm{keV}$) into that calculated for larger thermal WDM particle masses, $m_\mathrm{th,new}$, found in Ref.~\cite{Vogel:2022odl}, we use a linear fit relation,
\begin{equation}
\label{eq:th_correction}
    m_\mathrm{th,new} \approx 0.83 \, m_\mathrm{th,old}+0.36\, ,
\end{equation}
where the particle masses are in units of keV. This relation should be used to properly map any work which utilizes the previous WDM transfer function fit, in order to convert it into a more broadly calibrated thermal WDM particle mass value. We repeat our fitting process across the full parameter space and derive contours to describe the thermal WDM equivalent masses in order to apply signals and constraints directly to the SF parameter space. This is shown in Sec.~\ref{sec:3} and Fig.~\ref{fig:thermal_mass_plot}.

\section{\label{sec:3}Results}

Our primary results are shown in Fig.~\ref{fig:thermal_mass_plot}, which was calculated by using the {\tt GZA} code as described in the previous section over the full parameter space shown. The colored contour plot shows the equivalent thermal WDM particle mass for that position in parameter space. The white dot-dashed line at $m_\mathrm{thermal} = 4.1\,\mathrm{keV}$ is the best-fit value for the evidence for WDM from the Lyman-$\alpha$ forest \cite{Villasenor:2022aiy}, and the full parameter space is consistent with the $1\sigma$ inferred thermal WDM particle mass from that work. In Fig.~\ref{fig:thermal_mass_plot}, the red dashed line represents the current strongest bound on thermal WDM particle mass $m_\mathrm{th} > 8.5\,\mathrm{keV}$ (95\% CL) from a combination of galaxy counts \cite{Nadler:2021dft} and strong lensing (Zelko+) \cite{Zelko:2022tgf}. The gray exclusion region is formed from the strongest x-ray bound at each mass, resulting from \textit{NuSTAR} and \textit{SPI}/\textit{INTEGRAL} constraints \cite{Perez:2016tcq, Roach:2019ctw, Roach:2022lgo, Krivonos:2024yvm, Fischer:2022pse}.  

The contours for constant $L$ shift down with increasing $m_s$ because a smaller number density is needed to match the observed dark matter density, and less $L$-driven resonant conversion is needed. We find that the available window for SF to account for the entirety of the dark matter is excluded at much greater than $95\%\,\mathrm{CL}$ for $L<0.1$; however, extending to higher $L$ demonstrates the potential for a much larger parameter space available based on the current large values possible for the lepton number in the early Universe.  The shapes for the contours of constant \( m_{\mathrm{th}} \) have a tilt up and to the left due to the fact, as shown in Fig.~\ref{fig:production}, that higher $L$ in this region of the parameter space leads to a population of higher $q$ momenta modes, and effectively warmer dark matter. Note that at higher mixing angles, in the regions shown that are largely ruled out by x-ray observations, the contours reverse as the SF mechanism produces lower average $q$ distributions, as shown in Ref.~\cite{Cherry:2017dwu}.

As part of this work, we also developed power-law fitting functions for the thermal WDM particle mass and the lepton number contours in the relevant region of parameter space. These fits provide a compact and accurate representation of the numerical results, facilitating interpolation and analytical estimates without rerunning the full sterile neutrino production code. The thermal WDM particle mass \( m_{\mathrm{th}} \) is fit as a function of the sterile neutrino particle mass \( m_s \) and mixing angle \( \sin^2{(2\theta)} \), while the lepton asymmetry \( L \) required to form all of dark matter is similarly expressed in terms of the same parameters. The functional forms of the fits are:
\begin{align}
m_{\mathrm{th}} &= A\, m_s^a \, \left[\sin^2(2\theta)\right]^b, \quad 
\begin{cases}
A = 1.329 \\
a = 1.044 \\
b = 0.055\, ,
\end{cases} \\
L &= C \, m_s^\alpha \, \left[\sin^2(2\theta)\right]^\beta, \quad
\begin{cases}
C = 1.865 \times 10^{-12} \\
\alpha = -1.813 \\
\beta = -1.042\, ,
\end{cases}
\end{align}
where particle masses are in units of keV. 
Importantly, $m_\mathrm{th}$ in these relations is the more accurate thermal WDM particle mass calculated in Ref.~\cite{Vogel:2022odl}.

These expressions capture the dominant scaling behavior
in the high-lepton asymmetry regime and are validated to
remain within percentage-level error bounds across the range \( 15~\mathrm{keV} \leq m_s \leq 60~\mathrm{keV} \) and \( 10^{-15} \leq \sin^2(2\theta) \leq 3\times 10^{-13} \). Specifically, the average error for the thermal mass fit was 1.3\% while the average error for the lepton number fit was 8.3\%.

Interestingly, if one adopts the Zelko+ bounds, the remaining unconstrained parameter space requires an \( L\gtrsim 0.5 \). Note that the full parameter space in Fig.~\ref{fig:thermal_mass_plot} is preferred by Lyman-$\alpha$ observations and within the derived $1\sigma$ thermal WDM particle mass region given by $m_\mathrm{th} = 4.1^{+33.3}_{-1.5}\,\mathrm{keV}$ \cite{Villasenor:2022aiy}.

\section{\label{sec:4}Discussion and Conclusions}

In this paper, we examine the Shi-Fuller sterile neutrino dark matter production mechanism given the newly uncovered possibility of large lepton numbers in the early Universe, during the dark matter production era. In this context, we study the combined constraints from structure formation and x-ray astronomy. Interestingly, there are both structure formation constraints on WDM from lensing and galaxy counts, as well as a current preference from recent interpretations of Lyman-$\alpha$ forest measurements. Specifically, the thermal WDM particle mass preferred by an interpretation of Lyman-$\alpha$ forest observations is $m_\mathrm{th} = 4.1^{+33.3}_{-1.5}\,\mathrm{keV}$ \cite{Villasenor:2022aiy}, while constraints on WDM from galaxy counts and lensing are at the level of $m_\mathrm{th} > 8.4\,\mathrm{keV}$ (95\% CL) \cite{Nadler:2021dft} and $m_\mathrm{th} > 8.5\,\mathrm{keV}$ (95\% CL) \cite{Zelko:2022tgf}.\footnote{As discussed in Sec.~\ref{sec:level1}, we use the updated, more accurate thermal WDM transfer functions for these signals and constraints.}  These structure formation constraints and preferences lie in the parameter space for SF sterile neutrino dark matter production that are not constrained by x-ray astronomy. 
We find that the available parameter space for the SF mechanism is no longer closed, but opens up for $L\gtrsim 0.5$, $m_s\gtrsim 35\,\mathrm{keV}$, and $\sin^2 (2\theta) \lesssim 10^{-14}$ when adopting both structure formation and x-ray constraints.

There are several other observational and experimental probes of the SF mechanism for the production of sterile neutrino dark matter. One is a potential detection of $N_\mathrm{eff} > 3.044$ with high significance from upcoming CMB experiments \cite{CMB-S4:2016ple,SimonsObservatory:2025wwn}. This could be a signature of high-$L$ in the very early Universe \cite{Froustey:2024mgf}. $N_\mathrm{eff} \approx 3.5 $ is already  favored by the Hubble tension \cite{Escudero:2022rbq, Escudero:2024uea}. As discussed in Sec.~\ref{sec:lepton}, the primordial abundances are also very sensitive to nonzero lepton number (chemical potential), particularly in the electron-flavor neutrinos. Of interest specifically is the recent EMPRESS measurement of $Y_p$  that may indicate nonzero $L_e$ at the BBN epoch~\cite{Matsumoto:2022tlr,Burns:2022hkq,Escudero:2022okz,Yanagisawa:2025mgx}, which would expand the possibilities for larger $L_\alpha$ in all flavors at early times \cite{Froustey:2024mgf}. CMB observations are also becoming highly sensitive to $Y_p$ and $N_\mathrm{eff}$ independently, allowing for tests of primordial nucleosynthesis \cite{ACT:2025tim,SPT-3G:2025bzu}.

Since the parameter space remaining is at lepton numbers of order unity, we have identified approximations that must be  examined in this high-$L$ regime of early Universe physics. In the fortunate event that a future x-ray observatory detects a signal in this small mixing angle region, our work shows what parameter spaces are of interest for structure formation and potential nuclear decay probes \cite{Friedrich:2020nze,Martoff:2021vxp,KATRIN:2022spi,Lee:2025txi}, as well as the necessity of carefully examining the full quantum kinetic behavior of SF sterile neutrino dark matter production at high $L$.

In addition to the resonance of the SF mechanism, enhancing the active neutrinos' scattering rate through nonstandard interactions (so-called NSI) can allow access to $\sin^2 (2\theta) \gtrsim 10^{-14}$ for sterile neutrino dark matter production via oscillation-based production \cite{DeGouvea:2019wpf,Kelly:2020pcy,Berryman:2022hds,An:2023mkf}; however, this NSI mechanism is beyond the scope of this present paper, and was recently examined in more detail in Ref.~\cite{Vogel:2025gan}.  In addition to enhancing methods of production involving neutrino oscillations, sterile neutrino dark matter can also be produced by mechanisms that are independent of mixing, such as those from parent particle decay to sterile neutrinos, including the ``keV miracle'' from Higgs boson decay \cite{Kusenko:2006rh}, a Grand-Unified-Theory-scale mechanism \cite{Kusenko:2010ik}, vector boson decay \cite{Shuve:2014doa}, from thermal production and subsequent dilution by new particles \cite{Patwardhan:2015kga}, from a parent sterile neutrino \cite{Fuller:2024noz}, or from new active-sterile interactions \cite{Dev:2025sah}, including via varying Yukawa couplings \cite{Jaramillo:2022mos}.

In conclusion, given the viability of the SF mechanism at $m_s \gtrsim 35\,\mathrm{keV}$, it is best probed by x-ray observatories in the energy scale of $E_\gamma = m_s/2 \sim 20 \,\mathrm{keV}$, as well as future probes of structure formation at the $m_\mathrm{th}\gtrsim 10\,\mathrm{keV}$ WDM scale, which may corroborate either the evidence for WDM from Villasenor et al., or augment the constraints on WDM. 

\textit{Note added.} --- Recently, a related paper appeared on arXiv \cite{Akita:2025txo}. In that paper, the authors find similar expansion of the parameter space using a new production code which incorporates nonaveraged neutrino oscillations and additional effects from large chemical potentials.

\acknowledgements 

C.M.V., K.N.A. and H.G.E. acknowledge useful conversations with Francis-Yan Cyr-Racine, George Fuller, Jay Krishnan, Harri Parkkinen, Michael Ryan, and M.\ Cristina Volpe. J.F. thanks Cyril Pitrou for many useful discussions. K.N.A. is partially supported by U.S. National Science Foundation (NSF) Theoretical Physics Program Grant No.\ PHY-2210283. K.N.A. acknowledges support of the Institut Henri Poincaré (UAR 839 CNRS-Sorbonne Université), and LabEx CARMIN (ANR-10-LABX-59-01) in hosting the ``Dark Matter and Neutrinos'' program, where many of these discussions took place. J.F. is supported by the Network for Neutrinos, Nuclear Astrophysics and Symmetries (N3AS), through the National Science Foundation Physics Frontier Center Award No. PHY-2020275.

\bibliography{SF}
\bibliographystyle{apsrev4-1}
\end{document}